# Tailoring Effective Exchange Interactions via Domain Walls in Coupled Heisenberg Rings

Vanita Srinivasa* and Jeremy Levy[†]
*Department of Physics and Astronomy, University of Pittsburgh*
*Pittsburgh, PA  15260   USA*

*e-mail: vas9@pitt.edu
[†]e-mail: jlevy@pitt.edu

The nature of the exchange coupling variation in an antiferromagnetic spin-1/2 system can be used to tailor its ground-state properties. In particular, dimerized Heisenberg rings containing domain walls have localized states which can serve as "flying spin qubits" when the domain walls are moved. We show theoretically that, when two of these rings are coupled, the movement of the domain walls leads to modulation of the effective exchange interaction between the qubits. Appropriately chosen configurations of domain walls can give rise to ferromagnetic effective exchange. We describe how these spin rings may be used as basic building blocks to construct quantum spin systems whose properties are tunable by virtue of the exchange variation within the rings.

## I. Introduction

The spin degree of freedom plays a key role in proposals for quantum information processing in solid state systems[1-5]. Basic physical entities present in solids, such as electrons and certain nuclei characterized by a spin quantum number $s = \frac{1}{2}$, are intrinsically two-level quantum systems and therefore serve as natural realizations of quantum bits (qubits) [1,2]. While a single $s = \frac{1}{2}$ spin is itself a qubit, it may also be regarded as a fundamental building block that, together with interactions between spins, can be used to construct larger quantum systems having properties that do not exist for the individual constituents. In this sense, the spins and their interactions may be regarded as naturally-existing elements of a quantum "toolkit" for the construction of "designer quantum materials" with a variety of features.

Among the types of quantum systems that may be constructed are single qubits formed from multiple spins interacting via Heisenberg exchange[6-11]. The advantages of such qubits include the ability to perform universal quantum computation without requiring time-dependent external magnetic fields[6] as well as with Heisenberg exchange as the sole physical interaction[7,8], where the latter approach requires qubits composed of a minimum of three spins. The ground-state doublet present in the spectra of uniform antiferromagnetic (AFM) Heisenberg chains of an odd number of spins can be used to define energetically stable "spin cluster qubits"[9,10], which are protected from decoherence by the presence of a finite energy gap above the qubit states[11].

In addition to the number of spins, the exchange profile describing the set of interactions among spins may be varied. Collections of spins with modulated exchange give rise not only to stable qubits, but also to systems capable of faithfully transporting quantum information[12,13]. In particular, it is possible to effectively construct a quantum field using a one-dimensional dimerized AFM Heisenberg spin-$\frac{1}{2}$ chain such that its topological excitations serve as qubits[13].

Introducing a domain wall which separates the two possible states of dimerization into the exchange profile produces a topologically stable logical qubit, whose spin density is localized at the domain wall. Movement of the domain wall within a large spin system allows the localized spin density to be propagated over arbitrary distances. While the qubit remains encoded in the spin-$\frac{1}{2}$ ground-state doublet of the entire system, the moving domain wall effectively changes the location from which the quantum information present in the form of nonzero spin density may be accessed, producing a "flying spin qubit" which is stable against local disorder in the exchange profile. This system therefore combines the stability properties of multi-spin qubits with the ability to transport the qubits within the very spin system in which they reside.

In order for quantum information processing to be possible with flying spin qubits, pairs of these qubits in their stationary form, which we refer to in the present work as domain wall qubits, must be able to interact in a manner such that they become entangled. Entangling operations are essential elements of the set of quantum gates required to achieve universal quantum computation[14]. In order to explore methods for entangling domain wall qubits, mechanisms by which they can be controllably coupled must first be understood. In this goal lies the motivation for the present work. Related work on methods for coupling qubits encoded in AFM molecular rings[15,16] has involved switching on (off) the effective pairwise ring couplings by selectively exciting (de-exciting) one of the rings in each pair to a state lying outside (inside) the qubit space. This scheme allows quantum gates to be performed by applying global fields to a chain of AFM rings alternating between two types, and qubit-qubit couplings can be switched off despite the existence of permanent spin-spin couplings. More recently, the advantages of applying this scheme to a chain of modulated AFM spin triangles have been discussed in the context of implementing quantum gates between molecular qubits[17], and control over the effective coupling between molecular qubits via chemical modification of the intermolecular link has been demonstrated[18]. A method of entangling qubits via coupling to a uniform Heisenberg spin chain has also been proposed [19], where it was noted that the sign of the effective qubit-chain coupling depends on the spin site within the chain to which the qubit is coupled.

Here, we show that the exchange profiles for a pair of coupled dimerized AFM Heisenberg rings of $s = \frac{1}{2}$ spins containing domain wall qubits provide a means of tailoring the effective exchange interaction between the qubits. In this method, the system remains within the space spanned by the product states formed from the ground states of the rings. Tuning of the effective exchange is achieved by varying the positions of the domain walls within the rings. In all cases considered in the present work, the effective qubit-qubit exchange is found to be isotropic. Certain configurations of domain walls give rise to ferromagnetic (FM) effective exchange, despite the AFM nature of the spin-spin coupling between the rings. These features allow Heisenberg rings containing domain wall qubits to serve as the building blocks of a new class of designer quantum materials, and we explore a few examples of such systems in the present work.

We first present basic features of an analytical model for a single AFM spin triangle containing a domain wall qubit. This framework is then used to determine expressions for the effective qubit-qubit exchange as a function of the domain wall positions within a pair of coupled AFM spin triangles. A direct connection between the spin density variation within the rings and the effective exchange between the qubits is demonstrated. Extension of the results to larger coupled-ring systems is illustrated through numerical calculations for a pair of domain wall qubits in coupled five-spin rings. Constructions of an effective three-qubit FM triangle,

AFM rings of nine qubits with variable dimerization and a domain wall, and an effective spin-1 chain formed from qubits with alternating FM and AFM effective exchange are then demonstrated, using spin-triangle domain wall qubits as the basic building blocks in each case. Finally, a possible physical realization of the spin-1 chain using quantum dots is described.

## II. Antiferromagnetic Heisenberg spin rings

The Hamiltonian describing a one-dimensional system of $n_c$ spin-$\frac{1}{2}$ objects coupled by nearest-neighbor Heisenberg (isotropic) exchange interactions is

$$H_{n_c} = \sum_{k=1}^{n_c} J_k \left( \mathbf{S}_k \cdot \mathbf{S}_{k+1} \right) \equiv \sum J_k \left( \mathbf{S}_k \cdot \mathbf{S}_{k+1} \right). \tag{1}$$

Here, $\mathbf{S}_k = \left( S_k^x, S_k^y, S_k^z \right)$ is the vector operator for the $k^{\text{th}}$ spin, and $\hbar = 1$. The constants $\{J_k\}$ represent the strengths of the nearest-neighbor spin-spin exchange interactions, whose AFM nature is incorporated into the model by assuming $J_k > 0$ for all $k$. Periodic boundary conditions are included in Eq. (1) by letting $k \pm n_c \equiv k$. In the absence of external magnetic fields, both the square of the total spin angular momentum operator $S^2 = \left( \sum \mathbf{S}_k \right)^2$ and the total z-component $S_Z = \sum S_k^z$ commute with $H_{n_c}$, so that state and operator representations can be defined for subspaces of definite $S_Z$ and/or $S^2$.

The nature of the ground state of $H_{n_c}$ depends on the form of the exchange profile, which is defined as the set of coupling constants $\{J_k\}$. To illustrate particular features of this dependence, we consider the smallest possible closed spin chain, which has $n_c = 3$ spins. From

Eq. (1), the Hamiltonian is $H_3 = \sum_{k=1}^{3} J_k \left( \mathbf{S}_k \cdot \mathbf{S}_{k+1} \right)$. The full Hilbert space for this Heisenberg spin triangle is spanned by $2^3 = 8$ states and consists of subspaces characterized by energy eigenstates with fixed total spin quantum numbers $S = \tfrac{3}{2}$ and $S = \tfrac{1}{2}$. The AFM spin-spin exchange gives rise to a ground state with the minimum possible total spin[9, 10], which is $S = \tfrac{1}{2}$ for odd $n_c$. The space of $S = \tfrac{1}{2}$ states can be divided into two subspaces $(S, S_Z) = \left( \tfrac{1}{2}, \pm \tfrac{1}{2} \right)$, each of which is two-dimensional and defines a pseudospin. We choose the particular exchange profile[13]

$$J_k = \tilde{J}_0 + \tilde{J}_1 \cos\left( \frac{2\pi}{3}(k-1) - \varphi \right). \qquad (2)$$

The first term of this parametrization describes uniform exchange $\tilde{J}_0$, and the second term is a sinusoidal modulation with an amplitude $\tilde{J}_1$ and a phase $\varphi$. Here, we choose $\tilde{J}_0 > 0$ and $\tilde{J}_1 > 0$. Eq. (2) represents a typical exchange profile for the smallest possible spin system ($n_c = 3$) which can contain a flying spin qubit. In this analytical model, a domain wall is centered around the position in the three-spin ring defined by the phase $\varphi$, and varying $\varphi$ corresponds to moving the spin qubit within the ring.

The spectrum of the spin triangle Hamiltonian $H_3$ is determined by the presence or absence of modulation in the exchange profile given in Eq. (2). For the case of uniform exchange, $\tilde{J}_1 = 0$, a set of eigenstates which spans the $(S, S_Z) = \left( \tfrac{1}{2}, \tfrac{1}{2} \right)$ subspace is

$$\left| \uparrow \pm \right\rangle \equiv \frac{1}{\sqrt{3}} \left( \left| 001 \right\rangle + e^{\pm 2\pi i/3} \left| 010 \right\rangle + e^{\pm 4\pi i/3} \left| 100 \right\rangle \right), \qquad (4)$$

where $|0\rangle \equiv |s=\frac{1}{2}, s_z=\frac{1}{2}\rangle$ and $|1\rangle \equiv |s=\frac{1}{2}, s_z=-\frac{1}{2}\rangle$ are the single-spin basis states associated with the $z$ component of spin. A corresponding set of eigenstates for the $(S, S_Z) = (\frac{1}{2}, -\frac{1}{2})$ subspace is found by flipping all spins in Eq. (4):

$$|\downarrow \pm\rangle \equiv \frac{1}{\sqrt{3}}\left(|110\rangle + e^{\pm 2\pi i/3}|101\rangle + e^{\pm 4\pi i/3}|011\rangle\right). \tag{5}$$

The four states in Eqs. (4) and (5) are degenerate ground states for $\tilde{J}_1 = 0$. Introducing modulation into the exchange profile results in a splitting of the energies of the pseudospin states within each subspace of fixed $S_Z$ (Fig. 1). The $S = \frac{1}{2}$ energy eigenstates become $|\uparrow g_\varphi\rangle, |\uparrow e_\varphi\rangle, |\downarrow g_\varphi\rangle$, and $|\downarrow e_\varphi\rangle$, where $|\sigma g_\varphi\rangle = \frac{1}{\sqrt{2}}\left(|\sigma +\rangle - e^{i\varphi}|\sigma -\rangle\right)$ and $|\sigma e_\varphi\rangle = \frac{1}{\sqrt{2}}\left(|\sigma +\rangle + e^{i\varphi}|\sigma -\rangle\right)$ for $\sigma = \uparrow, \downarrow$. The sets of eigenstates for the case $\tilde{J}_1 \neq 0$ are therefore linear combinations of those in Eqs. (4) and (5), with the particular superpositions determined by the phase of the modulation $\varphi$. The degenerate ground states of the spin triangle are $|\uparrow g_\varphi\rangle$ and $|\downarrow g_\varphi\rangle$, which have energy $E_g = -3(\tilde{J}_0 + \tilde{J}_1)/4$ and are separated from the first excited states $|\uparrow e_\varphi\rangle$ and $|\downarrow e_\varphi\rangle$ with energy $E_e = -3(\tilde{J}_0 - \tilde{J}_1)/4$ by a gap $\Delta \equiv E_e - E_g = 3\tilde{J}_1/2$. Note that both the energies $E_g, E_e$ and the gap $\Delta$ are independent of $\varphi$ [13]. Within the ground-state subspace, the spin triangle can be regarded as a two-level system and serves as a single qubit[9, 10]. The finite gap present for all $\varphi$ serves to protect this qubit from decoherence[8, 11].

### III. Analytical model for effective exchange

We now consider a system of two coupled spin triangles with modulated nearest-neighbor exchange interactions (Fig. 2). The exchange within each of the triangles is assumed to be given

by the profile in Eq. (2). The Hamiltonian for this six-spin system can be written as

$H = H_0 + H'_{ij}$, where

$$H_0 = \sum_{k=1}^{3}\left\{\left[\tilde{J}_0 + \tilde{J}_1 \cos\left(\frac{2\pi}{3}(k-1) - \varphi_a\right)\right]\mathbf{S}_{ka}\cdot\mathbf{S}_{(k+1)a} + \left[\tilde{J}_0 + \tilde{J}_1 \cos\left(\frac{2\pi}{3}(k-1) - \varphi_b\right)\right]\mathbf{S}_{kb}\cdot\mathbf{S}_{(k+1)b}\right\} \quad (6)$$

describes the coupling within the triangles (labeled $a$ and $b$), and $H'_{ij} = J_r \mathbf{S}_{ia}\cdot\mathbf{S}_{jb}$ with $J_r > 0$ denotes the inter-triangle AFM spin-spin coupling. Fig. 2 illustrates the particular cases $H'_{ij} = H'_{33}$ and $H'_{ij} = H'_{21}$. The full Hilbert space for the system is spanned by the set of all possible product states of individual-spin-triangle basis states. For $J_r = 0$ (uncoupled triangles), the gap between the ground state $E_0 = 2E_g$ and the first excited state $E_1 = 2E_e$ is $E_1 - E_0 = 2\Delta$. In the limit $J_r \ll \Delta$, the coupling between the rings $H'_{ij}$ can be regarded as a perturbation relative to $H_0$[20-22], and the pair of triangles can be described within the subspace spanned by the product states of the spin-triangle ground states. To simplify the notation, we define $|\uparrow(\varphi)\rangle = |\uparrow g_\varphi\rangle$ and $|\downarrow(\varphi)\rangle = |\downarrow g_\varphi\rangle$. The ground-state product basis can then be written as

$$\left\{|\uparrow(\varphi_a)\rangle|\uparrow(\varphi_b)\rangle,\ |\uparrow(\varphi_a)\rangle|\downarrow(\varphi_b)\rangle,\ |\downarrow(\varphi_a)\rangle|\uparrow(\varphi_b)\rangle,\ |\downarrow(\varphi_a)\rangle|\downarrow(\varphi_b)\rangle\right\}. \quad (7)$$

The states in Eq. (7) are product states of the instantaneous ground states associated with the individual triangles for the set of domain wall phases $\{\varphi_a, \varphi_b\}$. In a system in which the domain walls are in motion, choosing these states as a basis is similar to choosing a reference frame which rotates with the domain wall positions. Here, we use the basis (7) to analytically describe the variation in the *static* qubit states as a function of $\varphi_a$ and $\varphi_b$.

Within the subspace defined by the states (7), the Hamiltonian $H_0$ in Eq. (6) is proportional to the identity operator $\mathbf{1}$ [10], and the full Hamiltonian can be rewritten as

$$H_3^{\text{eff}} = -\frac{3}{2}(\tilde{J}_0 + \tilde{J}_1)\mathbf{1} + J_{\text{eff}}(J_r, \varphi_a, \varphi_b)\mathbf{S}_a^{\Delta} \cdot \mathbf{S}_b^{\Delta}. \tag{8}$$

Here, $\mathbf{S}_a^{\Delta} = \sum_{k=1}^{3} \mathbf{S}_{ka}$ and $\mathbf{S}_b^{\Delta} = \sum_{k=1}^{3} \mathbf{S}_{kb}$ are the total spin operators for the individual triangles. The effective exchange $J_{\text{eff}}$ is a function of the strength of the coupling between the triangles $J_r$ as well as of the domain wall phases $\varphi_a$ and $\varphi_b$. Note that the nontrivial exchange term in Eq. (8) arises entirely from $H'_{ij}$, and also that the exchange interaction remains isotropic within the subspace, as was found for spin cluster qubits[9, 10]. The form of $J_{\text{eff}}(J_r, \varphi_a, \varphi_b)$ depends on the spins $ia$ and $jb$ involved in the coupling between the rings $H'_{ij} = J_r \mathbf{S}_{ia} \cdot \mathbf{S}_{jb}$. For $H'_{33} = J_r \mathbf{S}_{3a} \cdot \mathbf{S}_{3b}$ [Fig. 2(a)], the effective exchange is given by

$$J_{\text{eff}}^{33}(J_r, \varphi_a, \varphi_b) = \frac{J_r}{9}(1 + 2\cos\varphi_a)(1 + 2\cos\varphi_b), \tag{9}$$

while for $H'_{21} = J_r \mathbf{S}_{2a} \cdot \mathbf{S}_{1b}$ [Fig. 2(b)], the form of the effective exchange is

$$J_{\text{eff}}^{21}(J_r, \varphi_a, \varphi_b) = \frac{J_r}{9}\left(\cos\varphi_a - 1 + \sqrt{3}\sin\varphi_a\right)\left(\cos\varphi_b - 1 - \sqrt{3}\sin\varphi_b\right). \tag{10}$$

The origin of the exchange coupling variation in Eqs. (9) and (10) is directly related to the variation of the spin density at the sites $ia$ and $jb$. We show this by deriving Eq. (8) for $H'_{ij} = H'_{33}$. To do so, we calculate the matrix elements of $H'_{33}$ in the basis given in Eq. (7). The inter-triangle interaction term can be rewritten as

$H'_{33} = J_r \mathbf{S}_{3a} \cdot \mathbf{S}_{3b} = J_r\left[S_{3a}^z S_{3b}^z + \frac{1}{2}\left(S_{3a}^+ S_{3b}^- + S_{3a}^- S_{3b}^+\right)\right]$, where $S_{k\lambda}^{\pm} = S_{k\lambda}^x \pm i S_{k\lambda}^y$ for $\lambda = a, b$. The first

term of this expression has only diagonal nonzero elements, and the second term has only off-diagonal nonzero elements[10]. Using the fact that $\langle\downarrow(\varphi_\lambda)|S_{k\lambda}^z|\downarrow(\varphi_\lambda)\rangle = -\langle\uparrow(\varphi_\lambda)|S_{k\lambda}^z|\uparrow(\varphi_\lambda)\rangle$ and the representation defined by the order of the states in Eq. (7), we find

$$H'_{33} \to h_{11}\begin{pmatrix} 1 & 0 & 0 & 0 \\ 0 & -1 & 2 & 0 \\ 0 & 2 & -1 & 0 \\ 0 & 0 & 0 & 1 \end{pmatrix}, \qquad (11)$$

where $h_{11} \equiv J_r\langle\uparrow(\varphi_a)|S_{3a}^z|\uparrow(\varphi_a)\rangle\langle\uparrow(\varphi_b)|S_{3b}^z|\uparrow(\varphi_b)\rangle = J_r(1+2\cos\varphi_a)(1+2\cos\varphi_b)/36$.

Setting the matrix in Eq. (11) equal to that for a Heisenberg exchange interaction between two spin-$\frac{1}{2}$ objects in the standard basis $\{|00\rangle,|01\rangle,|10\rangle,|11\rangle\}$ gives $H'_{33} \to J_{eff}^{33}(J_r,\varphi_a,\varphi_b)\mathbf{S}_a^\Delta\cdot\mathbf{S}_b^\Delta$, which is the second term in Eq. (8) for the case $H'_{ij} = H'_{33}$. Here, $J_{eff}^{33} = 4h_{11}$, which agrees with Eq. (9) and may also be written as

$$J_{eff}^{33} = 4J_r\langle\uparrow(\varphi_a)|S_{3a}^z|\uparrow(\varphi_a)\rangle\langle\uparrow(\varphi_b)|S_{3b}^z|\uparrow(\varphi_b)\rangle. \qquad (12)$$

The quantities $\langle\uparrow(\varphi_a)|S_{3a}^z|\uparrow(\varphi_a)\rangle$ and $\langle\uparrow(\varphi_b)|S_{3b}^z|\uparrow(\varphi_b)\rangle$ are none other than the values of the spin densities at sites $3a$ and $3b$. We therefore find the result that the effective exchange between the spin-triangle qubits is equal to the product of the spin densities at the sites participating in the inter-triangle spin-spin coupling.

In particular, if these two spin densities are of opposite signs, the effective exchange is negative. With the convention chosen in the present work that positive values of the exchange are AFM, the negative sign corresponds to FM effective exchange. Note that this is true despite the AFM nature of the spin-spin coupling $J_r$ between the triangles. Because the spin densities vary with the phases $\{\varphi_a, \varphi_b\}$ of the domain walls within the triangles, these phases provide a method of controlling the effective coupling between the qubits. In other words, the *inter*-triangle coupling can be tuned via the *intra*-triangle coupling, and in particular, FM qubit-qubit coupling can in principle be realized with only AFM spin-spin coupling.

As an example, the variation of the effective exchange given by Eq. (9) for the coupled triangle pair in Fig. 2(a) is plotted in Fig. 3(a) as a function of $\varphi_a$, with $J_r = 1$ and $\varphi_b = 0$. This variation is independent of the values of $\tilde{J}_0$ and $\tilde{J}_1$. Note that as the domain wall in ring $a$ is moved around the ring, the effective exchange changes from AFM to FM and back to AFM, which reflects the changing spin density at site $3a$. A maximum in the FM exchange strength occurs for $\{\varphi_a = \pi, \varphi_b = 0\}$, while for the domain wall configurations $\{\varphi_a = 2\pi/3, \varphi_b = 0\}$ and $\{\varphi_a = 4\pi/3, \varphi_b = 0\}$, the qubit-qubit coupling is effectively zero. The vanishing exchange can be understood by considering the spin density within the triangles. For both of these domain wall configurations, the spin at site $3a$ belongs to a relatively strongly coupled pair whose ground state is the $S = 0$ singlet state $\frac{1}{\sqrt{2}}(|01\rangle - |10\rangle)$ of two spins. The spin density at site $3a$ is therefore zero, which results in the vanishing of the effective coupling in Eq. (12). Fig. 3(b) shows the energies of the eight lowest states of the coupled-spin-triangle pair as a function of $\varphi_a$ for $\tilde{J}_0 = 1$, $\tilde{J}_1 = 1$, $J_r = 0.1$, and $\varphi_b = 0$. The effective exchange splitting is apparent in the energies of the lowest four states, three of which are triply degenerate and one of which is nondegenerate. The presence of a relatively large gap between these lowest two energy levels and higher states for all values of $\varphi_a$ confirms the validity of the effective exchange approximation.

The method of modifying the spin density within a Heisenberg ring in order to produce FM effective exchange which is described here is closely analogous to techniques that have been used to synthesize crystals of organic radicals with intermolecular FM exchange[23], which involve stacking radicals in orientations such that atoms with spin densities of opposite sign are neighboring each other. In these systems, the presence of the unpaired electron in a cyclic radical plays a role similar to the domain wall in the Heisenberg spin rings considered in the present work. In addition, the importance of the relative orientations of a pair of stacked radicals to the nature of the overall intermolecular effective exchange was shown to be related to the overlap of the atomic orbitals between the radicals[24], with large overlap corresponding to AFM exchange and small overlap to FM exchange. A similar but much less sophisticated relationship for the Heisenberg ring systems containing domain wall qubits can be obtained by calculating the overlap $\left|\langle\sigma(\varphi_a)|\sigma(\varphi_b)\rangle\right|=\left|\cos\left[(\varphi_a-\varphi_b)/2\right]\right|$ between the ground states of the two spin triangles, which can be regarded as orbital-like overlap by noting that moving the domain wall in each ring corresponds to changes within an orbital degree of freedom for each qubit[13]. For $\varphi_b=0$, it is seen that $\left|\langle\sigma(\varphi_a)|\sigma(0)\rangle\right|\geq\frac{1}{2}$ for $0\leq\varphi_a\leq 2\pi/3$ and $4\pi/3\leq\varphi_a\leq 2\pi$, which are the regions of AFM effective exchange [Fig. 3(a)] while $\left|\langle\sigma(\varphi_a)|\sigma(0)\rangle\right|\leq\frac{1}{2}$ for $2\pi/3\leq\varphi_a\leq 4\pi/3$, which corresponds to the range over which the exchange is FM. AFM (FM) effective exchange is therefore seen to occur for domain wall locations associated with larger (smaller) orbital-like overlap values.

**IV. Numerical studies of effective exchange**

The discussion of effective exchange between domain wall qubits has so far focused on a particular analytical model, in which the exchange profile is given by Eq. (2). To illustrate the generality of the results, we consider alternative forms of exchange profiles which give rise to domain walls in dimerized Heisenberg rings defined by Eq. (1) with $n_c = 5$ and periodic boundary conditions. Numerical calculations are carried out in order to determine the effective exchange between the qubits, assuming that they are encoded in the ground states of the rings.

For each five-spin ring, we initially consider an exchange profile formed from two exchange constants $J$ and $aJ$, with $0 \leq a \leq 1$, in which there is a domain wall separating the two possible states of dimerization. General features of spin systems with such exchange profiles are discussed in Ref. 13. We consider a pair of spin rings (labeled $a$ and $b$) which each have the exchange profile $\{J_1 = J_4 = J, J_2 = J_3 = J_5 = aJ\}$ and are coupled by AFM exchange $J_r > 0$ [Fig. 4(a)]. This system has domain walls centered at sites $3a$ and $3b$. The ground-state doublet of a single AFM five-spin ring has $(S, S_z) = (\frac{1}{2}, \pm\frac{1}{2})$. These two states define a qubit[9, 10, 13] and are determined by numerical diagonalization of the Hamiltonian in Eq. (1) for $n_c = 5$. The spin densities of the two ground states are plotted as a function of site $k$ in Fig. 4(b) and show the spin density of the qubit localized around the domain wall at $k = 3$. Denoting the product basis constructed from the spin-ring ground states by $\{|\uparrow\rangle_a|\uparrow\rangle_b, |\uparrow\rangle_a|\downarrow\rangle_b, |\downarrow\rangle_a|\uparrow\rangle_b, |\downarrow\rangle_a|\downarrow\rangle_b\}$, where $\uparrow$ refers to the $S_z = +\frac{1}{2}$ state and $\downarrow$ to the $S_z = -\frac{1}{2}$ state, we determine the effective Hamiltonian within the subspace spanned by these states for $J = 1$, $a = 0.2$, and $J_r = 0.1$. For the coupling depicted in Fig. 4(a) (where both domain walls are at the positions defined to be zero) we find the effective Hamiltonian $-3.037\,\mathbf{1} + 0.096\,\mathbf{S}_a \cdot \mathbf{S}_b \equiv H_5^{eff}(0)$, where $\mathbf{S}_a$ and $\mathbf{S}_b$ are the total spin operators of rings $a$ and $b$, respectively. As was found for the coupled spin triangles and for spin cluster qubits[9, 10], the qubit-qubit exchange Hamiltonian is of the isotropic Heisenberg form (up to a term proportional to the identity, which simply corresponds to a uniform shift of all energies).

We now displace the domain wall within ring $a$ by $s$ sites in the direction of increasing site index with respect to the labeling in Fig. 4(a) by applying a discrete translation operator to $|\uparrow\rangle_a$ and $|\downarrow\rangle_a$, and we calculate the effective Hamiltonian $H_5^{eff}(s)$ in the shifted product basis for each distinct position of the domain wall. For the system in Fig. 4(a), there are three distinct positions, as seen by noting that the exchange profile is periodic with a period of $n_c = 5$ sites, and further that the sites $1a$ and $5a$ are equivalent by symmetry, as are the sites $2a$ and $4a$, so that $H_5^{eff}(1) = H_5^{eff}(4)$ and $H_5^{eff}(2) = H_5^{eff}(3)$. We find $H_5^{eff}(s) = -3.037\,\mathbf{1} + J_5^{eff}(s)\mathbf{S}_a \cdot \mathbf{S}_b$, where the values of $J_5^{eff}(s)$ are plotted in Fig. 4(c). Note the variation in the sign of $J_5^{eff}(s)$, indicating that the qubit-qubit exchange can be either AFM or FM, depending on the position of the domain wall within the five-spin ring. As in the case of the spin-triangle pair, the effective exchange variation results directly from the change in the spin density at site $3a$ as the domain wall in ring $a$ is displaced.

The relationship between the spin density within one of the $n_c = 5$ spin rings and the effective exchange between the qubits can be made more apparent by moving the domain wall more continuously. For each ring, we therefore choose the particular exchange profile

$$J_k = J(1+a)/2 + [J(1-a)/2](-1)^k \alpha(k-k_0), \tag{13}$$

where the staggered order parameter $\alpha(k-k_0) = (1/N)\sum_{r=-n}^{n}(-1)^r \tanh\left(\left[(k-k_0)-rn_c\right]/w\right)$

with $k_0 = (n_c+1)/2 + \Delta k$ denoting the position of the domain wall and

$N = \sum_{r=-n}^{n}(-1)^r \tanh\left(\left[n_c/2 - rn_c\right]/w\right)$. Exchange profiles of this form can be used to produce

flying spin qubits[13]. For the present case ($n_c = 5$, $k_0 = 3 + \Delta k$), we choose the parameter values

$J = 1$, $a = 0.1$, $w = 2$, and $n = 50$, along with $J_r = 0.1$ for the spin-spin coupling between the

rings. The effective exchange is determined numerically by diagonalizing the Hamiltonian for

the coupled pair of rings within the ground-state product subspace

$\left\{\left|\uparrow(\Delta k_a)\right\rangle\left|\uparrow(\Delta k_b)\right\rangle, \left|\uparrow(\Delta k_a)\right\rangle\left|\downarrow(\Delta k_b)\right\rangle, \left|\downarrow(\Delta k_a)\right\rangle\left|\uparrow(\Delta k_b)\right\rangle, \left|\downarrow(\Delta k_a)\right\rangle\left|\downarrow(\Delta k_b)\right\rangle\right\}$ and calculating

the energy gap between the singlet state (energy $E_s$) and the triplet states (energy $E_t$) of the two

qubits associated with the rings in the above basis. The exchange splitting

$J_5^{eff}(\Delta k_a, \Delta k_b = 0) = E_t(\Delta k_a, \Delta k_b = 0) - E_s(\Delta k_a, \Delta k_b = 0)$ is shown in Fig. 5(a) as a function of

$\Delta k_a$, where $J_5^{eff} > 0$ corresponds to AFM exchange and $J_5^{eff} < 0$ to FM exchange. Note that

$J_5^{eff}$ is periodic for two full revolutions of the domain wall around the ring, which arises from the

fact that the staggered order parameter $\alpha$ itself has a period of $2n_c = 10$ sites. Fig. 5(b) shows the variation of the spin density at site $3a$ for the state $\left|\uparrow(\Delta k_a)\right\rangle$ over the same range of values

of $\Delta k_a$. We find that a relation analogous to Eq. (12) holds:

$$J_5^{eff} = 4J_r \left\langle\uparrow(\Delta k_a)\right|S_{3a}^z\left|\uparrow(\Delta k_a)\right\rangle\left\langle\uparrow(0)\right|S_{3b}^z\left|\uparrow(0)\right\rangle. \tag{14}$$

Applying Eq. (14) to the spin density values in Fig. 5(b) with $J_r = 0.1$ and

$\left\langle\uparrow(0)\right|S_{3b}^z\left|\uparrow(0)\right\rangle = \left\langle\uparrow(0)\right|S_{3a}^z\left|\uparrow(0)\right\rangle$ reproduces exactly Fig. 5(a). We also find that the ratio

$J_r / \Delta E$, where $\Delta E$ denotes the gap between the qubit states and the next excited state for a single ring, is small for all $(\Delta k_a, \Delta k_b)$, its maximum value being $(J_r / \Delta E)_{max} \approx 0.153$. The effective exchange approximation therefore remains valid for the case of the coupled five-spin ring system.

## V. Construction of quantum spin systems by tailoring of the effective exchange

We have shown that both the magnitude and the sign of the effective exchange between the qubits encoded in the ground states of two dimerized AFM Heisenberg rings containing domain walls can be tailored by suitable modification of the spin density within the rings. This ability to tune the nature of the exchange allows the spin rings to serve as building blocks for a wide variety of quantum spin systems. Here we demonstrate some examples of systems which can be constructed by virtue of this method. We use spin triangles of the type discussed in Secs. II and III as the basic building blocks in order to simplify the analytical description of the exchange profile and the resulting spin density variation within each ring.

### A. Ferromagnetic triangle of qubits

As a first illustration, we consider the construction of an effective ferromagnetically-coupled triangle of *qubits* using only AFM spin-spin couplings. This system requires $n_\Delta = 3$ triangles with modulated exchange. The signs of the spin density at each site of a spin triangle having the exchange profile in Eq. (2) with $\varphi = \pi / 3$ are indicated in Fig. 6(a). Here, we show the signs of the spin density for the ground state $|\uparrow(\varphi)\rangle$ of the spin triangle, but one can equally well consider the spin density for $|\downarrow(\varphi)\rangle$. In the latter case, all spin density signs would simply be reversed. According to Eq. (12), it is the product of the spin densities at the sites involved in the spin-spin coupling between a pair of triangles that determines the sign of the effective exchange, which is independent of the pseudospin space (↑ or ↓) chosen for the individual qubits (provided that the same space is chosen for all of them). One possibility for achieving FM effective exchange between two qubits is to couple a site labeled "2" in one triangle to a site labeled "1" in another triangle [Fig. 6(b)], so that the product of the spin density values at the sites involved in each inter-triangle spin-spin interaction is negative. The effective exchange for this case is given by Eq. (10). Since all three rings are of the type shown in Fig. 6(a), $\varphi_a = \varphi_b = \varphi_c = \pi/3$, and the effective exchange between each pair of rings is $J_{eff}^{21}(J_r, \frac{\pi}{3}, \frac{\pi}{3}) = -2J_r/9$. For $J_r > 0$, this results in a triangle of qubits with uniform FM effective exchange.

## B. Dimerized Heisenberg triangle ring with domain wall

More complex systems of exchange-coupled qubits can also be constructed using the method discussed in the present work. In particular, it is possible to create a dimerized AFM Heisenberg ring of *qubits* containing a domain wall with spin triangles. Here, we show how this is possible using the domain wall configurations and inter-triangle couplings for $n_\Delta = 9$ triangles (Fig. 7). The extreme case of a single isolated qubit and strongly coupled dimers is illustrated in Fig. 7(a). The basic spin triangle building block for this case is shown, with sites of zero and positive spin density for the state $|\uparrow(\varphi)\rangle$ indicated. The coupling between each pair of triangles $a$ and $b$ is given by $H'_{31} = J_r \mathbf{S}_{3a} \cdot \mathbf{S}_{1b}$, which leads to the effective exchange Hamiltonian in Eq. (8) within the space spanned by (7), where

$$J_{eff}(J_r, \varphi_a, \varphi_b) = J_{eff}^{31}(J_r, \varphi_a, \varphi_b) = \frac{J_r}{9}(1 + 2\cos\varphi_a)(1 - \cos\varphi_b + \sqrt{3}\sin\varphi_b). \tag{15}$$

The $n_\Delta = 9$ qubit system with dimerized effective exchange and a domain wall is created using triangles with domain wall phases that alternate between $\varphi_1 = 0 + \Delta\varphi$ and $\varphi_2 = 2\pi/3 - \Delta\varphi$, except where $\varphi_1$ appears twice in a row, which creates a domain wall in the *effective* exchange profile. Fig. 7(a) shows the $n_\Delta = 9$ triangle ring for $\Delta\varphi = 0$. By setting $\Delta\varphi = \pi/3$, uniform AFM exchange can be achieved between the qubits. This case is depicted in Fig. 7(b), along with the basic spin triangle building block and the spin density signs associated with the exchange profile of the triangle. We note here that, due to the asymmetry of the exchange within the triangle at the domain wall relative to its neighboring triangles, the two effective AFM couplings between the isolated qubit and each of its neighboring dimers do not increase in an identical way as $\Delta\varphi$ is varied from 0 to $\pi/3$. Nevertheless, both couplings increase monotonically to their identical values at $\Delta\varphi = \pi/3$.

### C. Effective spin-1 chain

We now demonstrate a possible construction of an effective spin-1 AFM Heisenberg chain. In general, a spin-1 chain can be formed from spin-$\frac{1}{2}$ objects with alternating FM and AFM exchange, in the limit where the FM exchange tends to infinity[25]. Using the method discussed in the present work, a spin-1 chain can be approximated by a chain of coupled triangle qubits in which the domain wall configurations produce alternating FM and AFM exchange [Fig. 8(a)]. In order to determine a suitable set of domain wall phases $\{\varphi_a, \varphi_b\}$, we let the effective exchange function alternate between $J_{eff}^{33}$ [Eq. (9)] and $J_{eff}^{21}$ [Eq. (10)] and assume $J_{eff}^{33}(J_r^{33}, \varphi_a, \varphi_b) < 0$ and $J_{eff}^{21}(J_r^{21}, \varphi_b, \varphi_a) > 0$. Assuming all spin-spin couplings are AFM, this leads to the conditions $(1 + 2\cos\varphi_a)(1 + 2\cos\varphi_b) < 0$ and

$\left(\cos\varphi_b - 1 + \sqrt{3}\sin\varphi_b\right)\left(\cos\varphi_a - 1 - \sqrt{3}\sin\varphi_a\right) > 0$. A set of domain wall phases which satisfies these inequalities is $\{\varphi_a = \frac{\pi}{3}, \varphi_b = \pi\}$, which leads to $J_{eff}^{33}\left(J_r^{33}, \frac{\pi}{3}, \pi\right) = -2J_r^{33}/9$ and $J_{eff}^{21}\left(J_r^{21}, \pi, \frac{\pi}{3}\right) = 4J_r^{21}/9$. With $J_r^{33} > 0$ and $J_r^{21} > 0$, the chosen domain wall phases produce the desired alternating FM and AFM effective exchange. In order for this system to closely approximate a spin-1 chain, one also requires the ratio $\left|J_{eff}^{33}/J_{eff}^{21}\right| = J_r^{33}/2J_r^{21}$ to be large, while ensuring that both $J_r^{33}$ and $J_r^{21}$ remain much smaller than the gap $\Delta$, which is proportional to the triangle modulation amplitude $\tilde{J}_1$. A schematic of the constructed system is shown in Fig. 8(b). Note that the variation of the spin density within the basic triangle building blocks is of the same form as that shown in Fig. 7(b).

To explore the properties of the approximate spin-1 chain constructed here, the Hamiltonian for a $n_\Delta = 4$ triangle chain with $\tilde{J}_0 = 10$, $\tilde{J}_1 = 8$, $\varphi_a = \pi/3$, $\varphi_b = \pi$, $J_r^{33} = 1$, $0.05 \leq J_r^{21} \leq 0.5$, and periodic boundary conditions was diagonalized using the Lanczos method[26]. Fig. 8(c) shows the gap above the ground state, $E_{gap}$, as a function of the effective exchange ratio $J_{eff}^{21}/J_{eff}^{33}$. Note that a finite energy gap above the nondegenerate ground state is present for this spin triangle qubit chain for all values of $J_{eff}^{21}/J_{eff}^{33}$ shown. This finding agrees qualitatively with that expected for a spin-1 chain from Haldane's conjecture[27, 28], which suggests that an AFM Heisenberg integer spin chain possesses a finite gap above the ground state. The lowest gap is well approximated by its value obtained from diagonalizing the effective Hamiltonian

$$H_\Delta^{eff} = -\frac{3n_\Delta}{4}\left(\tilde{J}_0 + \tilde{J}_1\right)\mathbf{1} + \sum_{m=1}^{n_\Delta/2}\left[J_{eff}^{33}\left(J_r^{33}, \varphi_a, \varphi_b\right)\mathbf{S}_{2m-1}^\Delta \cdot \mathbf{S}_{2m}^\Delta + J_{eff}^{21}\left(J_r^{21}, \varphi_b, \varphi_a\right)\mathbf{S}_{2m}^\Delta \cdot \mathbf{S}_{2m+1}^\Delta\right] \quad (16)$$

with $n_\Delta = 4$, where the assumed periodic boundary conditions imply that $m \pm n_\Delta \equiv m$ [Fig. 8(d)]. The error [Fig. 8(d)] is seen to decrease rapidly with the increasing ratio $J_r^{33}/J_r^{21}$ (corresponding to more weakly coupled effective spin-1 objects), and its relatively small value for all values of $J_r^{33}/J_r^{21}$ shown indicates that the approximation of the four-triangle (12-spin) system with AFM spin-spin couplings by four qubits coupled via the appropriate effective exchange interactions remains valid.

## VI. Physical implementation

In order to experimentally verify the effective exchange effects derived in the present work and attempt to construct spin systems such as those described in Sec. V, it is necessary to find physical realizations of AFM Heisenberg spin rings with the appropriate exchange profiles. Here, we describe a possible implementation of the effective spin-1 chain described in Sec. V C which involves an array of quantum dots, each containing a single electron. In the absence of external magnetic fields, AFM exchange interactions between electron spins in neighboring quantum dots are favored[29]. This suggests a potential advantage of the method discussed in the present work, since FM exchange between multi-spin qubits can be achieved using only AFM spin-spin couplings, without the additional magnetic fields that are required for FM coupling of single-electron-spin qubits in quantum dots. Another distinction between electron spin qubits in quantum dots and domain wall qubits arises in the context of a material in which spin-orbit coupling is present. Within a quantum dot array fabricated from such a material, the movement of the domain wall required to tailor the effective exchange by translating the spin density can in principle occur with negligible movement of the electrons contained in the quantum dots. As a result, the spin density can be transported without coupling to its spatial motion, which is not possible for the individual electrons themselves due to the spin-orbit interaction.

Fig. 9(a) shows a configuration of spin triangles, which is modified relative to that shown in Fig. 8(b) but retains the alternating FM and AFM effective exchange interactions required to construct the spin-1 chain from the triangle qubits, as can be deduced from the spin density signs. The quantum dot array for realizing the effective spin-1 chain is illustrated in Fig. 9(b). In this system, the required variation in the AFM electron spin exchange strengths can be achieved via the relative differences in the interdot separations. The creation of precisely-controlled arrays of Ge/Si quantum dots has also been experimentally demonstrated[30], which in principle allows for the implementation of spin systems with AFM spin-spin exchange couplings of varying strengths.

Molecular magnets provide another potential means of realizing some of the systems discussed in the present work[16, 31]. In these systems, which often can be described by nearest-neighbor AFM Heisenberg exchange Hamiltonians, it is possible to synthesize exchange interaction strengths to desired values. Additionally, it has recently been shown[31] that the coupling of a uniform external electric field to the chiral degree of freedom that exists for the spin triangle $Cu_3$ can modulate the exchange in a form equivalent to the exchange profile in Eq. (2).

The method of tailoring effective exchange interactions presented here may be of interest for quantum information processing, as it allows for controllable coupling between qubits that can additionally be converted to flying spin qubits by moving the domain walls, allowing for high fidelity transport, and that can also be localized or delocalized by changing the dimerization strength of the exchange profile[13]. One possible scheme for controllably coupling domain wall qubits within very large dimerized AFM Heisenberg rings of spins interacting over a relatively small region may be imagined as follows: moving the qubits as far apart as possible from the interaction region within their respective rings, and subsequently increasing (decreasing) the dimerization strength, results in smaller (larger) spin density in the region of interaction. Thus, when the spin density is delocalized, the effective coupling is "on", while localization of the spin density effectively turns the coupling "off" due to the approximately zero spin density within the interaction region. The possibility of carrying out quantum entangling operations between domain wall qubits which can give rise to universal quantum computing is the subject of current and future studies.

## VII. Conclusion

We have demonstrated that the effective exchange between qubits encoded in the ground states of dimerized AFM Heisenberg spin rings containing domain walls may be tailored via the spin-spin exchange variation within the rings. This method is based upon the principle that the effective exchange originates from the spin density distribution of the domain wall qubits. By employing this method, we have shown that domain wall qubits may be controllably coupled, and that these qubits can serve as the building blocks of a wide variety of designer quantum materials. Finally, we have suggested a possible scheme for realizing an effective spin-1 chain based on an array of single-electron quantum dots.

**Acknowledgements**

We are grateful to C. Stephen Hellberg for providing the LMSCS code used to carry out the numerical Lanczos diagonalization, and for helpful comments. We also thank Florian Meier for insightful discussions. This work was supported by a NDSEG Fellowship (VS) and the Department of Energy, Basic Energy Sciences (DE-FG02-07ER46421).

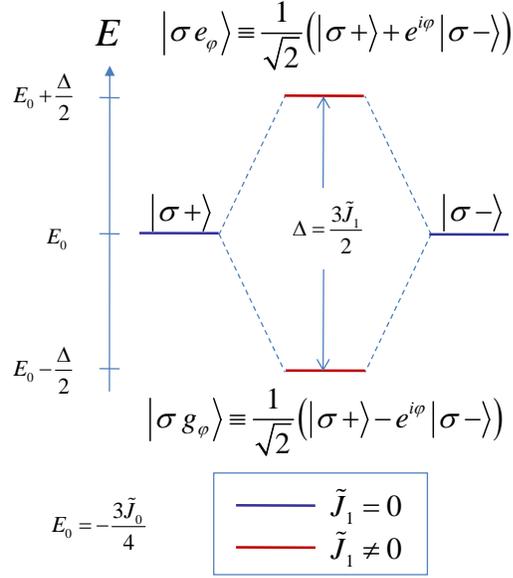

**Figure 1**: Energy level diagram illustrating the relationship between the energies of the pseudospin states $|\sigma\pm\rangle$ and those of the states $|\sigma g_\varphi\rangle$ and $|\sigma e_\varphi\rangle$ for the three-spin ring, with $\sigma = \uparrow, \downarrow$. The states $|\sigma\pm\rangle$ (blue) are eigenstates for the case of uniform coupling ($\tilde{J}_1 = 0$). Modulation of the exchange ($\tilde{J}_1 \neq 0$) gives the states $|\sigma g_\varphi\rangle$ and $|\sigma e_\varphi\rangle$ (red), which are separated by an energy gap $\Delta$.

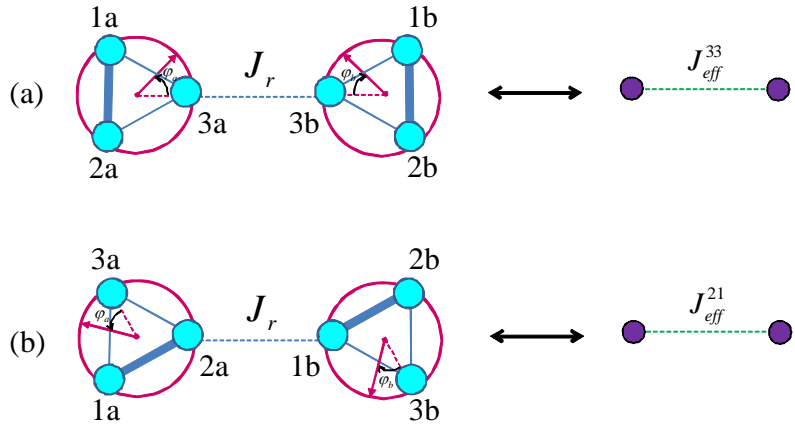

**Figure 2**: Pairs of spin triangles with exchange modulation parametrized by $\{\varphi_a, \varphi_b\}$ and coupled by antiferromagnetic exchange $J_r > 0$. The thicker and thinner lines drawn between the spins within the triangles indicate the stronger and weaker coupling strengths, respectively, for $\varphi_a = \varphi_b = 0$, which occurs when the domain walls are located at the site labeled "3" for each spin triangle. (a) Inter-triangle coupling $H'_{33}$. (b) Inter-triangle coupling $H'_{21}$. Within the ground-

state subspace, the triangle pairs in (a) and (b) are effectively pairs of qubits coupled by exchange $J^{33}_{eff}$ and $J^{21}_{eff}$, respectively [Eqs. (9) and (10)], as shown on the right.

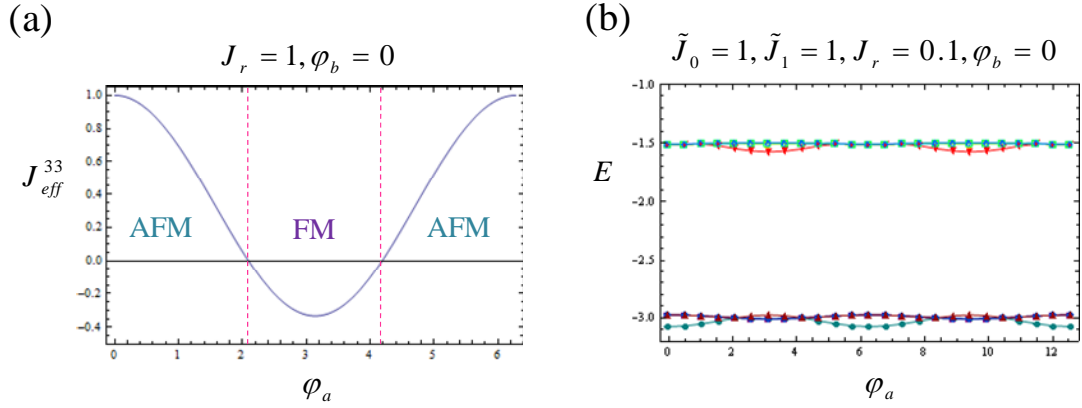

**Figure 3**: (a) Effective exchange $J^{33}_{eff}(J_r = 1, \varphi_a, \varphi_b = 0)$ given by Eq. (9) for the coupled spin triangle pair in Fig. 2(a), showing the ranges of domain wall locations $\varphi_a$ for which the exchange is antiferromagnetic (AFM) and ferromagnetic (FM). (b) Variation of the energies for the lowest eight states of the system in Fig. 2(a) as $\varphi_a$ is changed from 0 to $4\pi$ with $\tilde{J}_0 = 1$, $\tilde{J}_1 = 1$, $J_r = 0.1$, and $\varphi_b = 0$, showing the effective exchange splitting within the lowest two energies (four states, three of which are triply degenerate and one of which is nondegenerate) and a relatively large gap separating the lowest four states from higher levels for all values of $\varphi_a$.

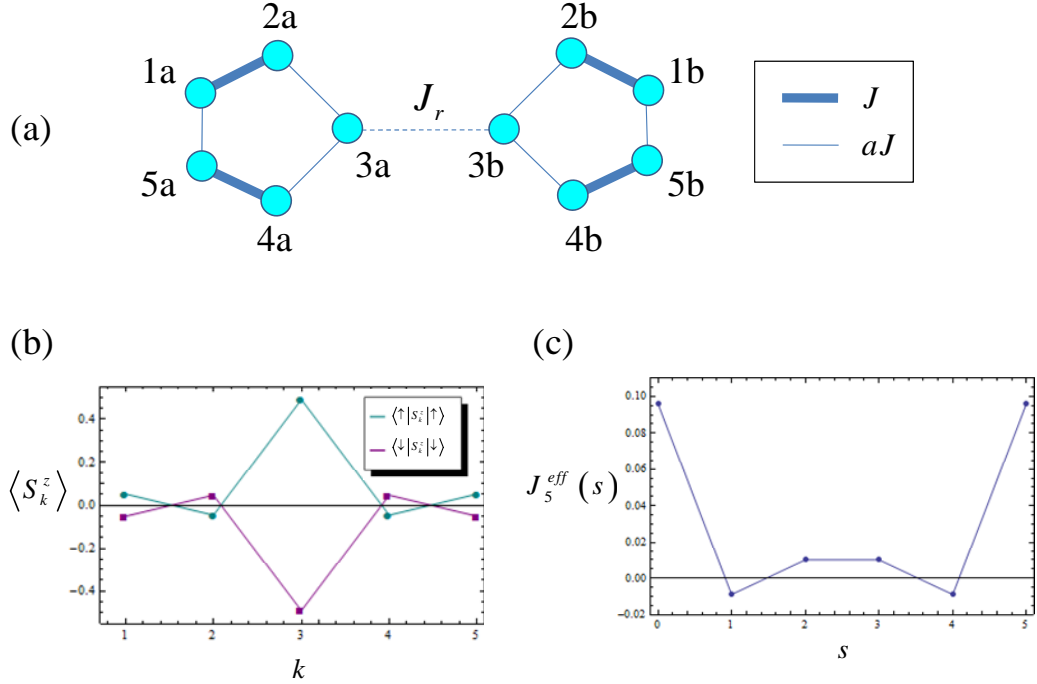

**Figure 4**: (a) Pair of $n_c = 5$ spin rings having dimerized exchange and containing domain wall qubits. The rings are coupled via AFM exchange $J_r > 0$, and the coupling strength between each pair of spins within the rings is either $J$ or $aJ$, with $0 \leq a \leq 1$. (b) Spin density of the $(S, S_z) = \left(\tfrac{1}{2}, \pm\tfrac{1}{2}\right)$ ground states of a single $n_c = 5$ spin ring of the type shown in (a) with $J = 1$ and $a = 0.2$, showing localization of the qubit around the position of the domain wall at $k = 3$. (c) Effective exchange $J_5^{eff}$ for $J_r = 0.1$ as a function of the number of sites $s$ by which the domain wall in ring $a$ shown in (a) is displaced, with $s = 0$ corresponding to the domain wall being located at site $3a$. The value of $s$ is defined to increase in the direction of increasing spin index ($3 \to 4 \to 5 \to 1 \to 2$).

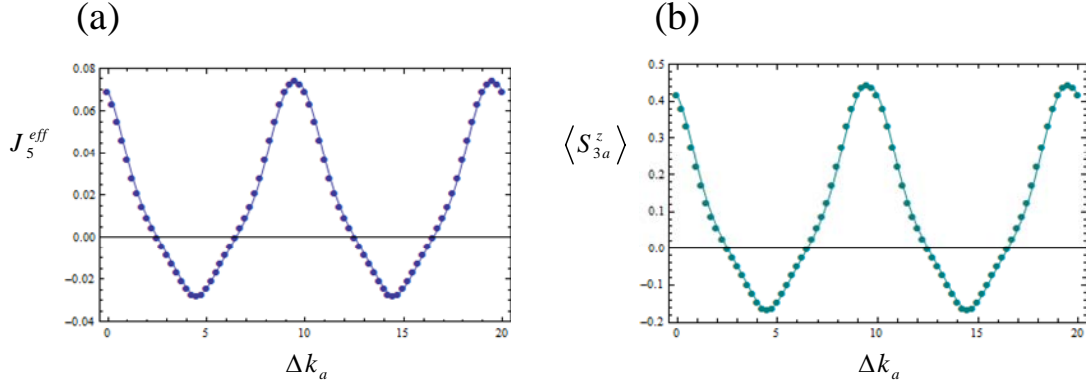

**Figure 5**: Effective exchange (a) and spin density variation at site $3a$ for the state $\left|\uparrow(\Delta k_a)\right\rangle$ (b), as a function of domain wall displacement $\Delta k_a$ for a pair of coupled $n_c = 5$ spin rings of the type shown in Fig. 4, with $J_r = 0.1$ and with the exchange profile for each ring given by Eq. (13), where $J = 1$, $a = 0.1$, $w = 2$, and $n = 50$.

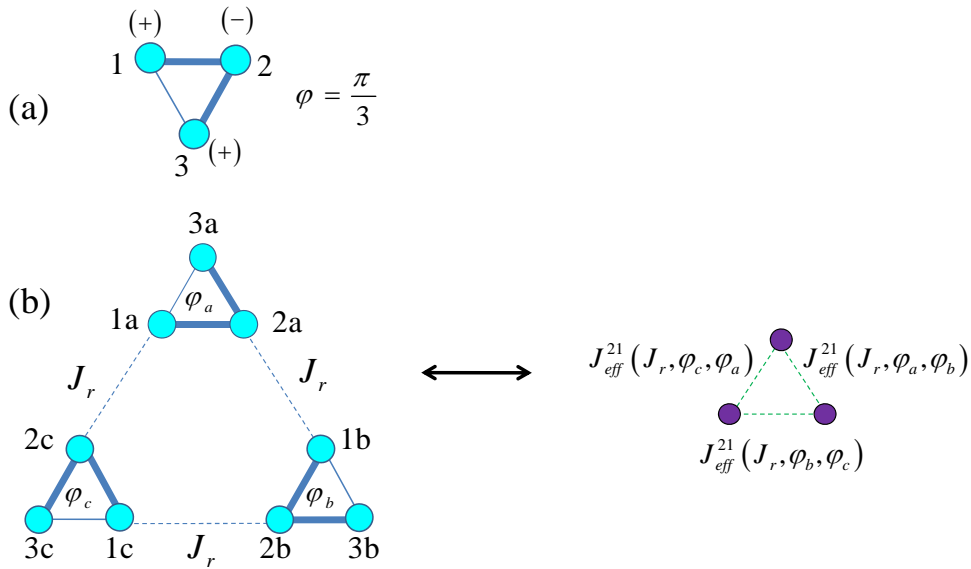

**Figure 6**: (a) Signs of spin density for the ground state $\left|\uparrow(\varphi)\right\rangle$ of a spin triangle having the exchange profile given in Eq. (2) with $\varphi = \pi/3$. (b) Possible coupling configuration for three spin triangles of the type in (a) which gives rise to an effective uniform FM qubit triangle (illustrated on the right). Here, $\varphi_a = \varphi_b = \varphi_c = \pi/3$.

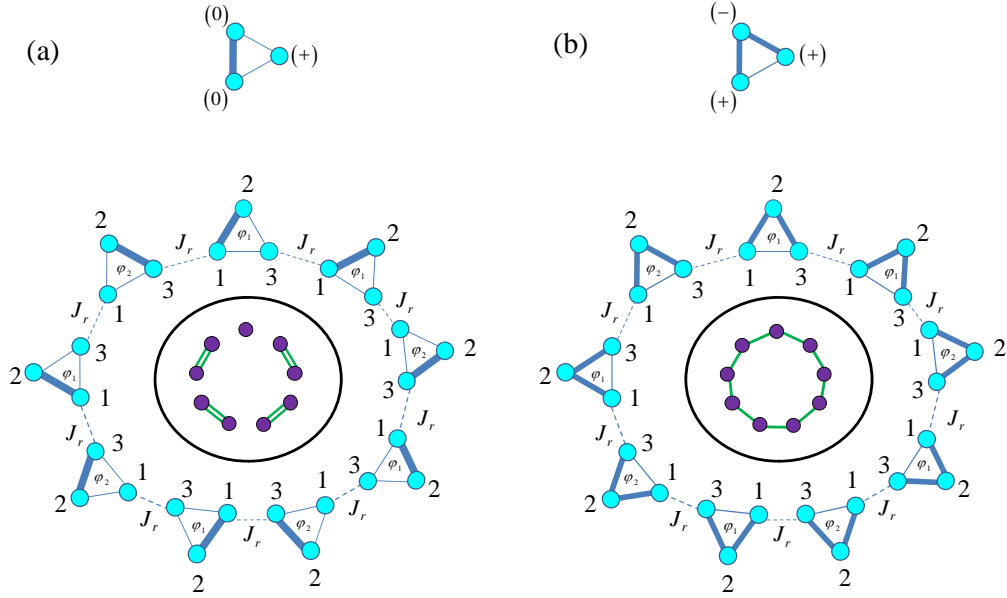

**Figure 7**: Effective AFM Heisenberg rings of $n_\Delta = 9$ spin triangle qubits with variable dimerization and a domain wall. The effective exchange between each pair of triangles is given by Eq. (15). (a) Basic spin triangle building block, indicating zero and positive spin density values within each qubit for the state $|\uparrow(\varphi)\rangle$, and constructed ring of coupled triangles giving rise to a single isolated qubit and strongly coupled dimers. Here, $\varphi_1 = 0$ and $\varphi_2 = 2\pi/3$. (b) Basic spin triangle building block, indicating positive and negative spin density values within each qubit for the state $|\uparrow(\varphi)\rangle$, and constructed ring of coupled triangles giving rise to a uniform Heisenberg ring of qubits. For this case, all domain wall phases are equal: $\varphi_1 = \varphi_2 = \pi/3$. Illustrations of the effective qubit systems for (a) and (b) are shown within the rings of triangles.

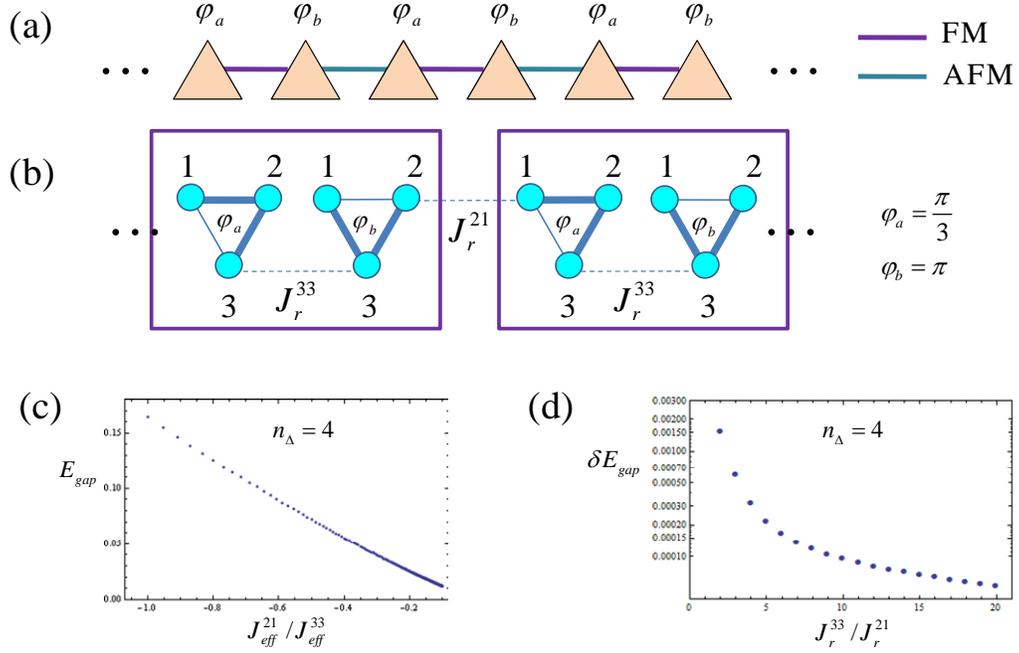

**Figure 8**: Effective spin-1 chain constructed from spin triangle qubits via tailoring of the effective exchange by domain walls. (a) Required effective system for a chain of coupled triangles, consisting of alternating domain wall phases $\varphi_a$ and $\varphi_b$ which produce alternating FM and AFM exchange in order to approximate a spin-1 chain. (b) Schematic of one possible solution for domain wall configurations and inter-triangle spin-spin coupling constants used to construct the effective spin-1 chain. The pairs of coupled triangles which act effectively as spin 1 objects are indicated by rectangles. (c) Lowest energy gap $E_{gap}$ as a function of the effective exchange ratio $J^{21}_{eff}/J^{33}_{eff}$ for a $n_\Delta = 4$ triangle chain of the type shown in (b) with $\tilde{J}_0 = 10$, $\tilde{J}_1 = 8$, $J^{33}_r = 1$, $0.05 \leq J^{21}_r \leq 0.5$, and periodic boundary conditions. (d) Difference $\delta E_{gap}$ between the value of the lowest energy gap for the full 12-spin system and that calculated by diagonalizing the effective Hamiltonian in Eq. (16), as a function of the ratio $J^{33}_r/J^{21}_r$.

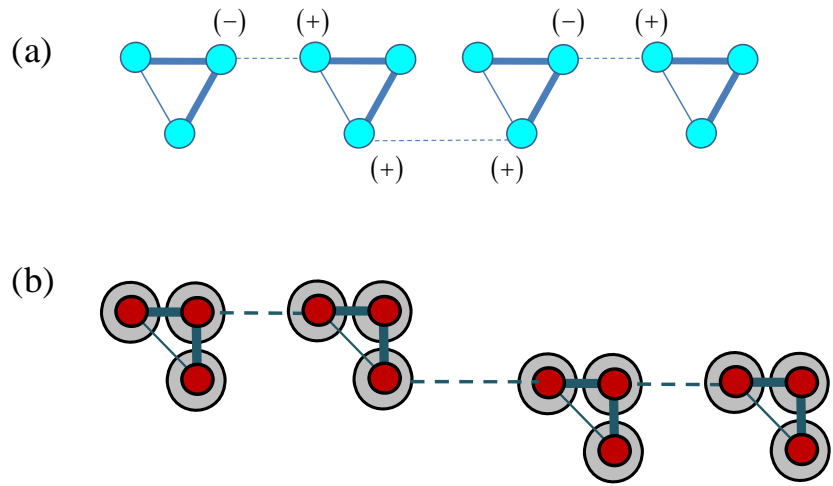

**Figure 9**: Possible physical realization of effective spin-1 chain. (a) Schematic diagram showing a coupling configuration for spin triangles which gives rise to alternating FM and AFM effective exchange. (b) Array of single-electron quantum dots for producing exchange coupling of the form shown in (a) in order to construct a spin-1 chain.